\documentstyle[aps]{revtex}

\begin{document}

\newcommand\beq{\begin{equation}}
\newcommand\eeq{\end{equation}}
\newcommand\bea{\begin{eqnarray}}
\newcommand\eea{\end{eqnarray}}

\newcommand\e{\epsilon}
\newcommand\gb{\bar{g}}
\newcommand\go{\delta g}
\newcommand\hh{\hat{H}}
\newcommand\ff{\stackrel{\leftrightarrow}{\Lambda}}
\def\ak#1#2{a_{#1}^{({#2})}}


\title{A two-dimensional interacting system obeying Fractional Exclusion 
Statistics}

\author{M. K. Srivastava$^{a,b}$, R. K. Bhaduri$^a$, J. Law$^c$, and M. V. N. 
Murthy$^d$}

\address
{$^a$Department of Physics and Astronomy, McMaster University,\\ 
Hamilton L8S 4M1, Ontario, Canada \\
$^b$Department of Physics, University of Roorkee, Roorkee 247667, India \\ 
$^c$ Department of Physics, University of Guelph,\\ 
Guelph N1G 2W1, Ontario, Canada\\
$^d$The Institute of Mathematical Sciences, Madras 600 113, India.}

\date{\today}

\maketitle

\begin{abstract}
We consider N fermions in a two-dimensional harmonic 
oscillator potential interacting with a very short-range repulsive 
pair-wise potential. The ground-state energy of this system is obtained  
by performing a Thomas-Fermi as well as a self-consistent Hartree-Fock 
calculation. The two results are shown to agree even for a small number 
of particles. We next use the finite temperature Thomas-Fermi method to 
demonstrate that in the local density approximation, these interacting 
fermions are equivalent to a system of noninteracting particles obeying 
the Haldane-Wu fractional exclusion statistics. 
It is also shown that mapping onto to a system of N noninteracting 
quasi-particles enables us to predict the energies of the 
excited states of the N-body system. 

\end{abstract} 
\vskip 1 true cm

\narrowtext
\section{Introduction}

Haldane~\cite{Haldane1} first proposed the generalised exclusion
statistics in the context of excitations in antiferromagnetic spin-chains. 
In this generalised scheme, one considers a finite dimensional
single-particle Hilbert space with $N$ identical particles. Increasing the
particle number by one results in the blocking off of $\alpha$
single-particle states for the others, the special cases $\alpha=1$ and
$\alpha=0$ corresponding to fermions and bosons respectively. It is to be
noted that the statistical parameter $\alpha$ is $N$-independent, and is
determined solely by the interaction strength between the particles.  An
exact realization of this statistics is found in the
Calogero-Sutherland-Moser (CSM) one-dimensional model~\cite{Calogero}, in
which the pair interaction potential falls off as the inverse square of
the relative straight-line distance between the particles, which are
either on an infinite line, or on a circle. Both the ground-state as well
as the excited state energies of this interacting CSM-system may be
exactly reproduced by a set of $N$ quasi-particles obeying Haldane
statistics~\cite{MS1}. The finite temperature distribution function for
the generalised exclusion statistics is known~\cite{Wu}. For the CSM
system, it has been shown~\cite{MS1} that the noninteracting
quasiparticles have such occupancies. In principle, generalised exclusion
statistics may be obeyed in higher dimensions also. 

The aim of this paper is to present a many-body Hamiltonian in two
dimensions that obeys fractional exclusion statistics. This model was
first proposed in~\cite{BMS}, and consisted of $N$ fermions in a
two-dimensional confinement potential, and interacting with a very
short-range repulsive two-body force.  For the one-dimensional CSM case,
the ground and the excited state energies of the many-body system are
analytically known, so one may test the mapping to the quasi-particle
spectrum. The difficulty in the two-dimensional case is that the energy
spectrum of the proposed many-body system is not known - in fact not even the
ground state energy is known, so the model cannot be tested unless
elaborate numerical calculations are performed. In ref\cite{BMS}, only the
Thomas-Fermi (TF) expression for the spatial density had suggested that
the system may obey fractional exclusion statistics. No attempt was made
in that paper to construct a quasi-particle scheme to map the energy
spectrum.  In this paper, we perform a self-consistent Hartree-Fock (HF)
calculation to estimate the gound-state energy of the two-dimensional
system.  It is found that the simple TF calculation for the energy agrees
surprisingly well with HF result. We then use the finite temperature TF
method to show that in the local-density approximation, the interacting
fermions are equivalent to a system of noninteracting particles obeying
the Haldane-Wu fractional exclusion statistics. Finally, we present a
quasi-particle spectrum that enables us to calculate the ground as well as
the excited state energies of the system. Although not expected to be
exact like the CSM in the one-dimensional case, the quasi-particle method
yields analogous finite-N corrections that lowers the energy.  Our model
may be useful for the description of interacting electrons in a quantum
dot, since it shows that the short-range component of the effective
interaction in a two-dimensional system may induce fractional exclusion
statistics.  The mapping to a system of non-interacting
quasi-particles further enables us to easily calculate the excited states
of a system of electrons in a quantum dot, albeit approximately. 
This, in the basis of electrons, would be a daunting task indeed.

\section{The Many-body Problem in Two-dimensions}

\subsection{The Hamiltonian}
We consider a system of $N$ fermions in two dimensions, confined in a 
one-body potential $V_1$, and interacting pair-wise through a short-range 
repulsive two-body interaction $V_2$. The Hamiltonian is given by 
\begin{equation}
H={1\over {2m}}\sum_{i=1}^N~p_i^2 + \sum_{i=1}^N~V_1(r_i) + \sum_{j<k}~
V_2(|{\bf r}_j-{\bf r}_k |)~.
\label{ham}
\end{equation}
As in \cite{BMS}, the spins are taken to be unpolarized. The mean-field 
expression for the energy at zero temperature is given by 
\begin{equation}
E[\rho({\bf r})]=\int d^2r \left[\tau({\bf r})+V_1(r)\rho({\bf r})+ 
{1\over 2}\{\rho({\bf r})\int~d^2r'
 \rho({\bf r'})V_2(|{\bf r}-{\bf r}'|)\newline
-{1\over 2}\int d^2r'|\rho({\bf r,r}')|^2 V_2(|{\bf r}-{\bf r}'|)\}\right]~,
\end{equation}
where $\tau({\bf r})$, $\rho({\bf r})$ are the spatial single-particle 
kinetic energy and number density, and $\rho({\bf r},{\bf r}')$ is the 
density-matrix. As shown in \cite{BMS}, the density-matrix expansion following 
Skyrme~\cite{Sky}, together with angle-averaging, yields the form  
\begin{equation}
E[\rho({\bf r})]=\int d^2r \left[\tau({\bf r})+V_1(r)\rho({\bf r})+ 
{1\over 4}\rho^2({\bf r}) M_0 + {1\over 8}\{\pi \rho^3({\bf r})-
[{\bf \nabla}\rho({\bf r})]^2\}M_2 +....\right],
\label{skyrme}
\end{equation}
where $M_n=\int d^2 r~ r^n V_2(r)$ is the $n^{th}$ moment of the two-body 
potential. If $V_2(r)$ has short range, the higher moments fall off 
rapidly. In this paper we only retain the lowest moment $M_0$, and drop 
the terms involving $M_2, M_3$ etc. This is tantamount to taking an 
effective interaction of zero range. In this approximation, we show that 
the interacting system may be mapped onto $N$ noninteracting quasi-particles 
obeying fractional exclusion statistics. We may therefore consider an 
effective Hamiltonian
\begin{equation}
\tilde{H}={1\over {2m}}\sum_{i=1}^N~p_i^2 + \sum_{i=1}^N~V_1(r_i) + 
{2\pi\hbar^2\over m}(\alpha-1)\sum_{j<k}~\delta(|{\bf r}_j-{\bf r}_k |)~.
\label{zero}
\end{equation} 
where $\alpha=(1+{mM_0\over {2\pi\hbar^2}})$ is dimensionless, and will 
play the role of the statistical parameter. Note from the above equation that 
$\alpha=1$ describes a system of noninteracting fermions.

\subsection{Thomas-Fermi (TF) approximation}
Consider, within the TF formalism, the above system at zero and finite 
temperatures. We show that this 
system of interacting fermions is equivalent to a set of noninteracting 
particles in $V_1({\bf r})$ obeying fractional exclusion statistics locally. 
The usage of the term {\it locally} will be explained shortly.
Let us first recapitulate the $T=0$ case. In the TF 
approximation, $\tau({\bf r})$ in Eq.(\ref{skyrme}) is replaced by 
$\hbar^2\pi \rho^2({\bf r})/2m$. For the zero-range interaction for $V_2$  
given by (\ref{zero}), only the lowest moment $M_0$ in Eq.(\ref{skyrme}) 
is nonzero. For this case, (with a spin-degeneracy factor of 2 included), 
the TF energy density functional is given by  
\begin{equation}
E[\rho({\bf r})]=\int d^2 r\left[{\hbar^2\over 2m}\pi\rho^2({\bf r})+
V_1({\bf r}) \rho({\bf r})+{\pi\hbar^2\over {2m}}(\alpha -1)\rho({\bf r})
\right].
\label{kumar}
\end{equation}
The ground state energy $E_0^{TF}$ and the spatial density $\rho_0^{TF}(r)$ 
are now determined by the variation $\delta(E-\mu(0) N)=0$, where $\mu(0)$ is 
the chemical potential at $T=0$, such that 
$\int d^2 r~ \rho_0^{TF}({\bf r})=N$. On perfoming the variation, we get 
\begin{eqnarray}
\rho_0^{TF}({\bf r})&=&{m\over {\pi \hbar^2 \alpha}}[\mu(0)-V_1({\bf r})],~~
{\bf r}\leq {\bf r}_0, \\
              &=&0~, {\bf r}>{\bf r}_0~,
\label{tf}
\end{eqnarray}
where ${\bf r}_0$ is the classical turning point, defined by 
$\mu(0)= V_1({\bf r}_0)$. 
The form of Eq.(\ref{tf}) for the spatial density prompted us to suggest in 
\cite{BMS} that it may also be generated by a set of noninteracting particles, 
but with a zero-temperature occupancy factor $1/\alpha$ instead of $1$. 
Indeed, the same feature was earlier found~\cite{SB} for CSM, where the 
ground-state energy was also reproduced correctly to leading order in $N$. 
The chemical potential itself is determined by 
the requirement that the spatial density, integrated over space, should 
give the particle number $N$. For the special choice $V_1(r)=m\omega^2 r^2/2$ 
which we will explore in some detail, this gives the relation 
\begin{equation}
\mu(0)=\hbar \omega \sqrt{\alpha N}.
\end{equation}  
For this case, so far as the chemical potential is concerned, the 
{\it energy scale has been stretched by a factor} $\sqrt{\alpha}$, an 
important clue for the construction of the quasi-particle spectrum later. 
On performing the $r-$integration in (\ref{kumar}) with the $\rho_0(r)$ 
given by (\ref{tf}), and making use of the above expression for $\mu$, 
we get the ground state energy 
\begin{equation}
E_0^{TF}={2\over 3}\alpha^{1/2}N^{3/2} \hbar\omega~.
\label{kume}
\end{equation}
Our considerations here are all based on the TF model. It is important to 
test the accuracy of the model, specially for small $N$, by using a more 
elaborate method. To this end, we present in the next section the 
self-consistent Hartree-Fock results.

Before doing this, however, we 
demonstrate that this system approximately obeys, within the TF frame-work, 
fractional exclusion statistics even at a finite temperature $T$. 
A system of noninteracting particles at a finite temperature 
obeying fractional exclusion statistics has the following  distribution 
function at equilibrium temperature $T$~\cite{Wu} 
\begin{equation}
n_{\alpha}(\epsilon,T)={1\over {w(\epsilon,T)+\alpha}}~,
\label{wu1}
\end{equation}
where $w$ is the solution of the equation $w^{\alpha}[1+w]^{1-\alpha}=
\e^{(\epsilon-\mu)/T}$. For the special case of a two-dimensional 
translationally invariant system (with constant spatial density $\rho$ and 
constant single-particle density of states), Wu~\cite{Wu} had shown that 
\begin{equation}
\mu(T)=\alpha {\pi\hbar^2\rho\over m} + T \ln [1-\exp(-\pi\hbar^2 \rho/mT)]~.
\label{wu2}
\end{equation}
In \cite{BMS}, we applied finite temperature TF for a system of fermions in a 
potential $V_1({\bf r})$, and interacting with a zero-range force. In this 
case it is easily shown that  
\begin{equation}
[\mu(T)-V_1({\bf r})]=\alpha {\pi\hbar^2\rho ({\bf r},T)\over m} + 
T \ln [1-\exp(-\pi\hbar^2 \rho ({\bf r},T)/mT)]~.
\label{wun1}
\end{equation}
Here $\rho({\bf r},T)$ is given by~\cite{BMS}
\begin{equation}
\rho({\bf r},T)={mT\over {\pi\hbar^2}}~
\ln \left[1+exp \{(\mu(T)-V({\bf r},T))/T\}\right]~.
\label{dip2}
\end{equation}
This is the {\it local} version of Eq.(\ref{wu2}) above, where 
$\mu(T)$ and a constant $\rho$ are replaced by 
their local values $[\mu(T)-V_1({\bf r})]$ and $\rho({\bf r},T)$ respectively. 
So, in the presence of a fixed  external potential and a small number $N$ 
of electrons, as in a quantum dot, for example, fractional exclusion 
statistics may be obeyed in the {\it local-density} approximation. The latter 
term means 
that at any point ${\bf r}$, the local fermi-energy and spatial density 
obey the same equations as their counterparts in the translationally 
invariant system. This generally is true in the interior of the system, 
but deviations may occur at the edges if the spatial density changes 
rapidly in a scale compared to the size of the system. The local-density 
approximation has been successfully used in 
sophisticated nuclear many-body calculations even with a relatively small 
number of nucleons~\cite{PR}, where a large fraction of them may reside 
in the surface region. 

In case the single-particle density of states is not a constant, the 
Wu-equation (\ref{wu2}) is generalised to the following form~\cite{SB}:
\begin{equation}
\mu(T)=\alpha \nu +T \ln [1-exp(-\nu/T)]~,
\label{dip1}
\end{equation}
where 
\begin{equation}
\nu(T)=\int_0^{\infty}~n_{\alpha}(\epsilon,T) d\epsilon~.  
\label{kaisa}
\end{equation}
We emphasize that Eq.(\ref{dip1}) is quite general, and does not make any 
restrictive assumption about the density of states. 
We now show that our interacting fermionic 
system also obeys a {\it local} version of Eq.(\ref{dip1}) in the TF 
frame-work. To demonstrate this, we define 
\begin{equation} 
\nu({\bf r},T)=\int_0^{\infty}~d\epsilon~n_F({\bf r,p},T)~,  
\label{raj1}
\end{equation}
where 
\begin{equation}
n_F({\bf r,p},T)={1\over {exp[(p^2/2m+V({\bf r})-\mu)/T]+1}}~.
\label{raj2}
\end{equation}
For the system given by the effective Hamiltonian (\ref{zero}), 
\begin{equation}
V({\bf r},T)=V_1({\bf r})+{\pi\hbar^2\over m}(\alpha-1)\rho({\bf r},T)~.
\label{dip3}
\end{equation} 
Replacing $d\epsilon$ in Eq.(\ref{raj1}) by $p~dp/m$ at a fixed point 
${\bf r}$, and performing the p-integration, 
we then obtain 
\begin{equation}
\nu({\bf r},T)={\pi\hbar^2\over m} \rho({\bf r},T)~.
\label{raj4}
\end{equation} 
Using Eqs (\ref{raj4}) and (\ref{wun1}), both based on the interacting fermion
system, we can then write
\begin{equation}
[\mu(T)-V_1({\bf r})]=\alpha \nu({\bf r},T) +T \ln [1-exp(-\nu({\bf r},T)/T)]~.
\label{dip5}
\end{equation}
This is the local version of Eq.(\ref{dip1}) that we set out to show, and 
emphasizes that the local density approximation is applicable even when 
the single-particle density of states is not a constant.

We now go on to check the accuracy of the TF approximation for the ground 
state by performing a self-consistent HF calculation in the next section.

\subsection{The Hartree-Fock (HF) calculation}   

The advantage of performing a HF calculation with $\tilde{H}$ given by 
(\ref{zero}) is that the Fock term is local. A little algebra shows that the 
HF single-particle orbitals $\psi_i({\bf r})$ with energies $e_i$ obey the 
one-body equation 
\begin{equation}
\left[\hat{h}_0+{\pi\hbar^2\over m}(\alpha-1)\rho({\bf r})\right] 
\psi_i({\bf r})=e_i \psi_i({\bf r})~,
\label{hf}
\end{equation}
where $\hat{h}_0=\hat{p}^2/2m+m\omega^2 r^2/2$, and 
$\rho=\sum_i^{'}\psi^{*}\psi$, 
the prime indicating that the sum is over the occupied orbitals only.
Note that the one-body potential in (\ref{hf}) is the confining harmonic
oscillator in $\hat{h}_0$, {\it plus} the contribution coming from the direct 
and the exchange matrix-elements of the two-body interaction. Let us 
denote the latter by 
\begin{equation}
U({\bf{r}})= {\pi\hbar^2\over m}(\alpha-1)\rho({\bf r})~.
\label{you}
\end{equation}
The potential $U$, proportional to $\rho$, is determined by the occupied single-particle 
orbitals. Therefore the equations have to be solved iteratively, 
starting with a trial $\rho$, until self-consistency is achieved to a 
specified accuracy. To this end, the orbitals $\psi_i({\bf r})$ are expressed 
in the basis set generated by the unpertubed harmonic oscillator. The basis 
set, of course, has to be truncated in practice. For numerical accuracy, we 
find that if $\cal{N}$ shells are occupied in the unperturbed configuration, 
then at least $2\cal{N}$ shells should be included in the basis set. 
For $N=12$ particles, where ${\cal N}=3$, we have tested that the 
self-consistent results remain essentially unaltered when the dimension of 
the basis set is increased from shells $6$ to $10$.
The HF single-particle Hamiltonian, given by the left-hand side of 
Eq.(\ref{hf}) 
(with a trial $\rho$), is represented by a matrix in this basis. It is 
diagonalsied to obtain the orbitals and their eigenvalues. The density $\rho$ 
and the corresponding HF Hamiltonian are recalculated, and the 
procedure repeated for self-consistency. Typically, for moderate strength 
of the interaction ($\alpha=3/2$), about $40$ iterations or more are needed 
to obtain self-consistent occupied orbitals $\psi_i$. 
The HF ground-state energy is given by 
\begin{equation}
E_0^{HF}=\sum^{\prime}_i\left(e_i-{1\over 2}<\psi_i|U|\psi_i> \right)~,
\label{hfe}
\end{equation}
where $U$ is defined by Eq.(\ref{you}), and the sum over $i$ is over occupied 
states only.
The self-consistent results for 3 different systems 
($N=12$, $20$ and $42$ fermions), each for 2 values of $\alpha$, are 
presented in Table 1. These $N$-values are closed shells configurations 
for $h_0$. For non-closed shell systems, we found that numerical convergence 
for self-consistency is harder to achieve. Interestingly, as is 
shown in Table 1, $E_0^{HF}$ is very close to $E_0^{TF}$ calculated from 
Eq.(\ref{kume}). In Fig.1 we display, for $N=20$ and $\alpha=3/2$, the shape 
of the self-consistent HF one-body potential, $(m\omega^2 r^2/2+U(r))$ as a 
function of the dimensionless variable $R=(m\omega/\hbar)^{1/2} r$, and the 
corresponding low-lying single-particle energies $e_i$ in units of 
$\hbar \omega$. For comparison, the unperturbed harmonic 
oscillator and its energy eigenvalues are also shown in the same diagram. 
Two features are to be noted : (a) the relatively large upward shift at the 
center in the HF one-body potential, resulting in an {\it overall} 
shift in the single-particle spectrum, and (b) the almost equal 
but reduced spacing between the levels in the HF spectrum, with the 
degeneracy between the $(2s,1d)$ states being intact to 4 significant 
figures, and the splitting in $(2p,1f)$ and $(3s,2d,1g)$ subshells remaining 
negligble. These characteristics suggest that, like the CSM, the ground 
and the excited states of the N-particle system are shifted by the same 
overall constant whose magnitude is determined by the stength of the 
interaction. 
(For systems in which the last filled shell is only partially occupied, 
preliminary results indicate that the splittings between the states within a 
shell are larger.) We also note that the states $(n,l)$ and $(n,-l)$ remain 
strictly degenerate. 
We also found that in every case, the spacing between the HF shells  
are almost equal, with an effective oscillaor constant given by 
$\omega/\sqrt{\alpha}$. This behaviour of the HF $e_i$'s is very different 
from the quasi-particle spectrum $\epsilon_i$'s presented in the next 
section, the latter actually spreading out by a factor $\sqrt{\alpha}$.  
We may express the HF single-particle energies in the form 
\begin{equation}
e_n=\Delta(\alpha,{\cal N}_0)+(n+1)\omega/\sqrt{\alpha}~,n=0,1,..
\end{equation}
where ${\cal N}_0$ is number of filled shells in the system. The gap 
$\Delta$ is proportional to $({\cal N}_0+1)$, and the constant of 
proportionality increases with $\alpha$.

As mentioned in the previous section, the TF result (\ref{tf}) had suggested 
that the ground state energy of the interacting system may be estimated 
by filling up each level of the corresponding {\it noninteracting} system by a 
fraction $1/\alpha$, but preserving the total particle number. We call this 
the ``global'' fractional-statistics rule. It is useful only for the 
ground state, in contrast to the quasi-particle formalism given in the next 
section. In Table 1, the numerical values for $E^G$(global) are also given 
for comparison.

\section{The Quasi-particle Spectrum}

We wish to express the energy of an interacting N-particle system as a sum of 
noninteracting quasi-particle energies, both for the ground and the excited 
states. This can be done {\it exactly} for CSM, which behaves as an 
{\it ideal} gas obeying fractional exclusion statistics.  For the two-
dmensional model presented here, we cannot prove that such is the case. 
Nevertheless, we show that when the degeneracies of the levels are taken 
into account, a consistent scheme for the quasi-particle spectrum may be 
formulated, obeying the same rules as given for 
CSM~\cite{MS1,MS2}. This procedure  reproduces the large-N TF results for the 
ground state, and predicts the finite-N corrections. 
Following the steps of ~\cite{MS1}, we then show that such a 
mapping leads naturally to the thermodynamic distribution function 
formulated in \cite{Wu}.   

In what follows, we take the two-dimensional harmonic oscillator as the 
confining potential, and a spin-degenaracy of two. At the outset, we clarify 
certain definitions 
that are used in this presentation. The unperturbed harmonic oscillator 
spectrum is specified by {\it shells}, containing degenerate {\it states} 
in the shell. For example, the $j^{th}$ shell is at energy $j\omega$ and 
has $\eta_j=2j$ states in it, where we have included the spin degeneracy of 
two in $\eta_j$. The quasi-particle spectrum has exactly the 
same structure of these shells and states, except that the energy of a  
quasi-particle is determined by the occupancies of the states below it, as 
specified by the rules below. In this mapping, the unperturbed energy 
$\epsilon_k=k\omega$ of a particle is mapped onto the quasi-particle energy 
$\epsilon_A(k,\alpha)$. The analogy with CSM leads us to suggest the 
following      
\begin{equation}
\epsilon_A(k,\alpha)=\epsilon_k-  \omega (1-\sqrt{\alpha}) {\cal N}(\epsilon_k,0),
\label{mk1}
\end{equation}
where ${\cal N}(\epsilon_k,0)$ is the sum of the occupancy fractions of 
the shells with energies $\epsilon < \epsilon_k$. This is given by 
\begin{equation}
{\cal N}(\epsilon_k)=\sum_{j=1}^{\infty}\Theta(\epsilon_k-\epsilon_j)
{1\over {\eta_j}}\sum_{i=1}^{\eta_j} n_{j,i}~,
\label{mk2}
\end{equation}
where $\Theta(x)=0$ for $x\leq 0$ and $1$ for $x> 0$, and $n_{j,i}$ is the 
occupancy ($0$ or $1$) of the $i^{th}$ state of the $j^{th}$ shell. 
An example is illustrated in Fig.2, where (a) is an unperturbed harmonic 
oscillator excited state configuration in which one particle from the 
$(2s,1d)$ shell has been excited to the next $(2p,1f)$ shell. In (b), 
the corresponding mapping to the quasi-particle spectrum is shown. We use 
the notation $\tilde{\omega}=\sqrt{\alpha}\omega$. Using Eq.(\ref{mk2}), 
we note that ${\cal N}(\epsilon_1)=0,~ {\cal N}(\epsilon_2)=1,~ 
{\cal N}(\epsilon_3)=2,~ {\cal N}(\epsilon_4)=17/6,~ 
{\cal N}(\epsilon_5)=71/24$. Now we may use Eq.(\ref{mk1}) to deduce the 
quasi-particle energies as shown in Fig. 2(b).

For the ground state of a closed-shell configuration, all the shells are 
fully filled upto (say) the ${\cal N}_0^{th}$ shell, and are 
unoccupied thereafter. The above rules then lead to   
\begin{eqnarray}
{\cal N}(\epsilon_j)&=&j-1,~~1\leq j\leq {\cal N}_0~; \nonumber\\
                      &=&{\cal N}_0,~~j>{\cal N}_0~.
\label{mk3}
\end{eqnarray}
The energy levels $\epsilon_A(k,\alpha)$ in the ground-state 
configuration are given by $\omega,\omega+\tilde{\omega},
\omega+ 2\tilde{\omega}, ...
\omega+({\cal N}_0-1) \tilde{\omega}, 2\omega+({\cal N}_0-1)\tilde{\omega}, 
3\omega+({\cal N}_0-1)\tilde{\omega}, ....$.
The ground state energy is then given by 
\begin{eqnarray}
E_0^{(qp)}&=&\sum_{k=1}^{{\cal N}_0} \epsilon_A(k,\alpha)\eta_k~,\nonumber\\
        &=&2\sum_{k=1}^{{\cal N}_0} k\epsilon_k-2\omega (1-\sqrt{\alpha})
           \sum_{k=1}^{{\cal N}_0} k(k-1)~, \nonumber\\
        &=&\omega \sqrt{\alpha} {\cal N}_0 ({\cal N}_0+1)(2{\cal N}_0+1)/3~
           -\omega (\sqrt{\alpha} - 1){\cal N}_0({\cal N}_0+1)~.
\label{mk4}
\end{eqnarray}
where ${\cal N}_0$ is the last filled level. Expressing ${\cal N}_0$ in 
terms of $N$, 
\begin{equation}
N=2\sum_{k=1}^{{\cal N}_0} k~,
\label{mk5}
\end{equation}
the ground state energy is given by 
\begin{equation}
E_0^{(qp)}={2\over 3}\omega \sqrt{\alpha}N^{3/2}\left[1+{1\over {8 N}}+...
\right] - (\sqrt{\alpha}-1)\omega N~.
\label{mk6}
\end{equation}
The term proportional to $N^{3/2}$ is the same as the TF expression 
(\ref{kume}), while the others 
are the finite-N corrections arising from the microscopic approach taken 
here. The values obtained for $E_0^{(qp)}$ by Eq.(\ref{mk6}) are 
displayed in Table 1. Note that these are always lower than the corresponding 
HF values, as is to be expected. Note also that for $\alpha=0$, Eq.(\ref{mk6})
 gives the bosonic limit with all particles in the lowest energy state at 
$\omega$. Although the expression (\ref{mk6}) is only valid for a 
closed-shell system, the ground state energy of an ``open shell''  
configuration may also be calculated easily using this formalism.

The excited state energies for {\it any} configuration 
$\{n_{k,i}\}$ may now be written using Eqs.(\ref{mk1}, \ref{mk2}):
\begin{eqnarray}
E[\{n_{k,i}\}]&=&\sum_{k=1}^{\infty}\epsilon_A(k,\alpha)~
                 \sum_{i=1}^{\eta_k} n_{k,i}~,\\
              &=&\sum_{k=1}^{\infty}\epsilon_k~\sum_{i=1}^{\eta_k}n_{k,i}-
                 \omega~(1-\sqrt{\alpha})\sum_{k'(\leq k)=1}^{\infty}
                 \sum_{k=1}^{\infty}{1\over {\eta_{k'}}}\sum_{i=1}^{\eta_k}
                 \sum_{i'=1}^{\eta_k'}n_{k,i}n_{k',i'}~.
\label{mk7}
\end{eqnarray}
Note that the energy levels $\epsilon_A(k,\alpha)$ depend on the configuration 
$\{n_{k,i}\}$. 
Like the CSM, the excitation energies of the system are given only by the 
first term in Eq.(\ref{mk7}). The second term depends only on the total 
number of 
particles. The correspondance with CSM and the agreement of the leading 
N-dependent term of $E_0^{qp}$ with the TF and the HF results  suggests 
that (\ref{mk7}) may give a good description for the excited states.  
As an aside we remark that if the effective interaction 
between $N$ electrons in a quantum dot is dominated by a short range 
interaction, then the above expression for the 
total energy as a sum of the quasiparticle energies immediately yields 
not only the approximate ground state but also its excited states. A 
direct calculation of these excited states is in general extremely 
difficult.

We now consider the distribution function for the quasi-particles. 
First consider the ground state $(T=0)$. So far as the energy {\it spacing} 
between the occupied quasi-particle levels are concerned, it is given by 
$\tilde{\omega}=\sqrt{\alpha} \omega$.   
The number of particles $\Delta N_A$ in an energy interval between 
$\epsilon_A$ and $\epsilon_A+\Delta\epsilon_A$ is given by 
\begin{eqnarray}
\Delta N_A &=& {1\over {\tilde{\omega}^2}}\epsilon_A 
               \Delta \epsilon_A~, \nonumber\\ 
           &=& {1\over \alpha}{1\over {\omega^2}}\epsilon_A\Delta\epsilon_A~.
\label{mk8}
\end{eqnarray}
Thus the occupancy factor for the filled quasi-particle states in 
the energy scale of $\omega$ is $1/\alpha$ instead of $1$. This is precisely 
the $T=0$ distribution function for the occupied states in fractional 
exclusion statistics, irrespective of dimensionality.

In conclusion, we have shown that for fermions in a two-dimensional 
harmonic  potential interacting with a very short-range repulsive 
two-body force, the TF calculation for the ground state energy agrees well 
with the corresponding HF result. The HF self-consistent single-particle 
states get almost a constant upward shift, while still retaining equal 
spacings in a slightly shallower potential. Gaining confidence with this  
success of the TF method, we apply it for the finite temperature problem 
in a confinement potential $V_1({\bf r})$. We find that in this approximation, 
the interacting fermions obey the same equations {\it locally} as to 
be expected from a system of noninteracting particles obeying fractional 
exclusion statistics. This is true for any confinement potential, so long 
as the finite-range effects of the two-body interaction is neglected. 
It means that for such a system, the interacting fermions in the bulk 
obey fractional exclusion statistics, atleast so far as energies are 
concerned. For the special case of harmonic confinement, a 
quasi-particle spectrum is constructed to predict the ground- as well as 
the excited state energies of the system. This is analogous to the CSM case 
in the one-dimensional problem.  

This research was supported by NSERC (Canada). One of the authors (MKS) 
would like to thank the Department of Physics and Astronomy for its 
hospitality. We would also like to thank R. Shankar for useful discussions.


\begin{table}
\caption{Ground state energy of N-particle (closed shell) systems for 
various $N$'s (column 1) and interaction strengths (column 2). 
Column 3 displays the sum of the occupied orbital 
energies, and the next the HF ground state energy (see Eq.(\ref{hfe}).   
Column 5 is the TF energy given by Eq.(\ref{kume}). Column 6, with the 
heading $E^G$, is the ``global'' approximation as explained at the end of 
the HF section. The last column is calculated from the ``microscopic'' 
quasi-particle approach, as given by Eq.(\ref{mk6}).}
\begin{tabular}{ccccccc}
\multicolumn{1}{c}{$N$} &
\multicolumn{1}{c}{$\alpha$} &
\multicolumn{1}{c}{$\Sigma '_i e_i$} &
\multicolumn{1}{c}{$E^{HF}$} &
\multicolumn{1}{c}{$E^{TF}$} &
\multicolumn{1}{c}{$E^{G} $} &
\multicolumn{1}{c}{$E^{qp}$} \\ 
\tableline 
 12   & 5/4   & 34.424 & 31.301 & 30.984 & 32.000 & 29.890 \\
 12   & 3/2   & 39.969 & 34.278 & 33.941 & 34.667 & 31.598 \\
 20   & 5/4   & 73.781 & 67.079 & 66.667 & 68.000 & 64.723 \\
 20   & 3/2   & 85.694 & 73.472 & 73.030 & 73.333 & 68.991 \\
 42   & 5/4   & 223.818 & 203.478 &202.879 &204.400 & 198.526 \\
 42   & 3/2   & 260.002 & 222.886 &222.243 &224.000 & 213.465 \\
\end{tabular}
\end{table}

\begin{figure} 

\caption{Plot of the HF self-consistent 
potential $(m \omega^2 r^2/2+U(r))$ and the corresponding 
single-particle energies $e_i$ (in bold), 
for $N=20$  and $\alpha=3/2$. For comparison, $V_1(r)=m\omega^2 r^2/2$ and 
its energy levels (in light) are also shown. The distance $R$ is 
in units of  ${\sqrt {\hbar/m\omega}}$, and 
the energy scale is in units of $\hbar\omega$.}\label{figure1}

\vspace{.25in}
\caption{The quasi-particle spectrum for a one-particle one-hole 
excited state configuration for $N=12$. The crosses denote 
the particles, and the 
circle the hole. On the left, the configuration 
is shown for the harmonic potential (unperturbed), to the right 
is the quasi-particle spectrum to 
which the interacting system has been mapped.}\label{figure2}
\end{figure}

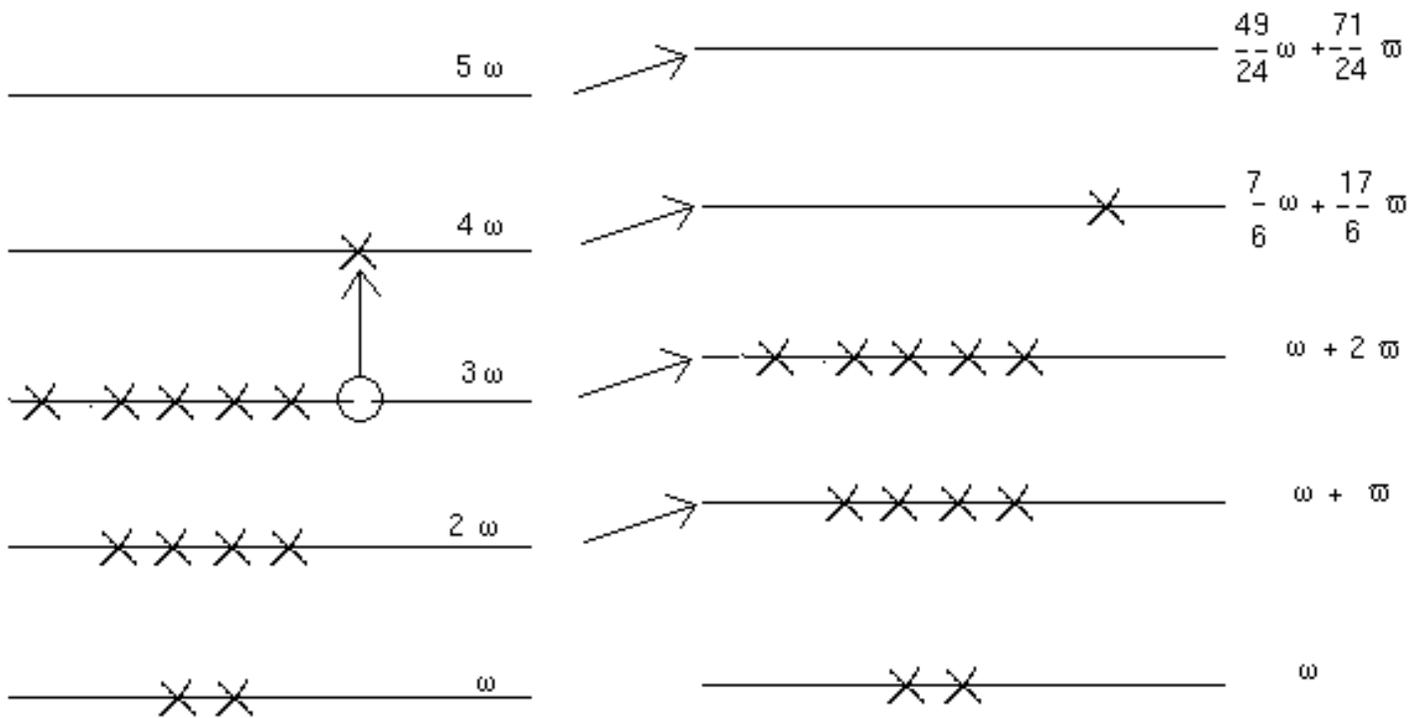
\end{document}